\documentclass[%
 reprint,
 amsmath,amssymb,
 aps,
]{revtex4-2}

\usepackage{graphicx}
\usepackage{dcolumn}
\usepackage{bm}

\usepackage{dsfont} 

\usepackage{mathrsfs}

\usepackage{booktabs}

\begin{document}

\preprint{APS/123-QED}

\title{Electron juggling: Approaching the atomic physics limit of the attempt rate in trapped ion photonic interconnects}

\author{I. D. Moore, B. M. White, B. Graner, and J. D. Siverns}
\affiliation{%
 IonQ, 3755 Monte Villa Pkwy, Bothell, WA 98021
}%

\date{\today}

\begin{abstract}
Photonic interconnects are a key technology for scaling up atomic based quantum computers. By facilitating the connection of multiple systems, high-performance modular quantum processing units may be constructed to perform deeper and more useful algorithms. Most previous implementations of photonic interconnects in trapped ions utilize the scheme of preparing a state, exciting it, and collecting single photons from decays of the excited state. State preparation is responsible for the vast majority of the total attempt time, often taking hundreds of nanoseconds to several microseconds. Here, we describe and analyze a novel technique called ``electron juggling" to speed up photonic interconnects by reducing the state preparation step substantially. Using a theoretical framework, we illustrate how this scheme can significantly increase remote entanglement generation rates, approaching the atomic physics limit of the attempt rate in trapped-ion photonic interconnects. Our results indicate that this scheme holds the possibility of achieving remote entanglement generation rates of over 1,000 Bell pairs per second.

\end{abstract}

\maketitle


\section{\label{sec:intro}Introduction}

Trapped ions are one of the most promising platforms for scalable quantum computing \cite{HAFFNER2008155,Egan2021}, simulation \cite{Blatt2012,Monroe2021} and networking \cite{Stephenson2020, Craddock2019, Walker2018, Hannegan2021C}. They exhibit long coherence times and some of the highest-fidelity operations of all qubit substrates \cite{Langer2005,Smith2025, Loschnauer2025, Wang2021,Hughes2025}. However, scaling up a trapped ion quantum processing unit (QPU) to a large number of qubits is non-trivial; as more ions are added, more control fields in the form of lasers, microwaves and electrode voltages will be required to manipulate and physically shuttle ions between zones of interest \cite{Kielpinski2002}. To avoid these increasing control field demands, ion chains can be made longer, which reduces shuttling overhead at the cost of motional-mode crowding and increased crosstalk between ions. Photonically linking QPUs with short chains of ions is a promising method to scale both quantum computers and networks while avoiding both of these deleterious effects \cite{Duan2010, Monroe2014, OReilly2024, Main2024,Inlek2017,Stephenson2020,Saha2024,Moehring2007,Maunz2009,Slodicka2013,Casabone2013,Hucul2015,Krutyanskiy2023}.

The remote entanglement generation (REG) rate of photonic interconnects is important for the performance of a photonically linked quantum computer. If the REG rate is too low, computation ions will decohere while they wait for sufficient Bell pairs to be generated for a remote gate. Increasing the remote entanglement generation rate beyond that of current records \cite{Stephenson2020,OReilly2024} is important to enable high-fidelity operations in photonically linked quantum computers running deep circuits \cite{Duan2010}. A high REG rate is even more critical where the fidelity of the ``raw" Bell pair is insufficient and entanglement distillation \cite{Bennett1996} is required, since then multiple Bell pairs will need to be generated to purify one acceptable pair.

The REG rate of a photonic interconnect channel can be increased by improving photon collection efficiency and/or increasing attempt rate. In the case of the former, work has been carried out to increase the collection efficiency of photons emitted from trapped ions using high numerical aperture (NA) lenses \cite{Shu2011, Ghadimi2017, Carter2024}, cavities \cite{Stute2012, Walker2021, Casabone2013, Krutyanskiy2019, Krutyanskiy2023} as well as frequency conversion of high-loss blue photons to infra-red and telecommunications bands to reduce losses over distances \cite{Zaske2012, Bock2018, Krutyanskiy2019, Hannegan2021C, Hannegan2022}. In the latter case, sympathetic cooling ions have been employed to avoid the need to pause interconnect procedures to perform cooling \cite{OReilly2024}. In typical approaches, state preparation is performed, followed by excitation and photon collection from decay of the excited state \cite{OReilly2024, Main2024, Inlek2017, Stephenson2020, Saha2024, Moehring2007, Maunz2009, Slodicka2013, Hucul2015}. The state-preparation step slows the REG rate down significantly. In this paper, we describe and simulate our ``\textit{electron juggling}" method, which skips state-preparation and results in an attempt rate limited instead by the lifetime of the excited state. 

Finally, it may seem that this method is only feasible in short-range quantum networks, since otherwise, as in long-range quantum networking, the attempt time will be dominated by the photon propagation time in the optical fiber. However, in long-range quantum networks, non-destructive measurements near the network nodes can be employed to flag the presence of a photon early on in the fiber \cite{Hannegan2021}. This shortens the average attempt time significantly, at which point ``electron juggling" may be incorporated into the long-range network for further speed improvements.


\section{\label{sec:electron juggling}Electron Juggling}

\begin{figure}
    \centering
    \includegraphics[width=1\linewidth]{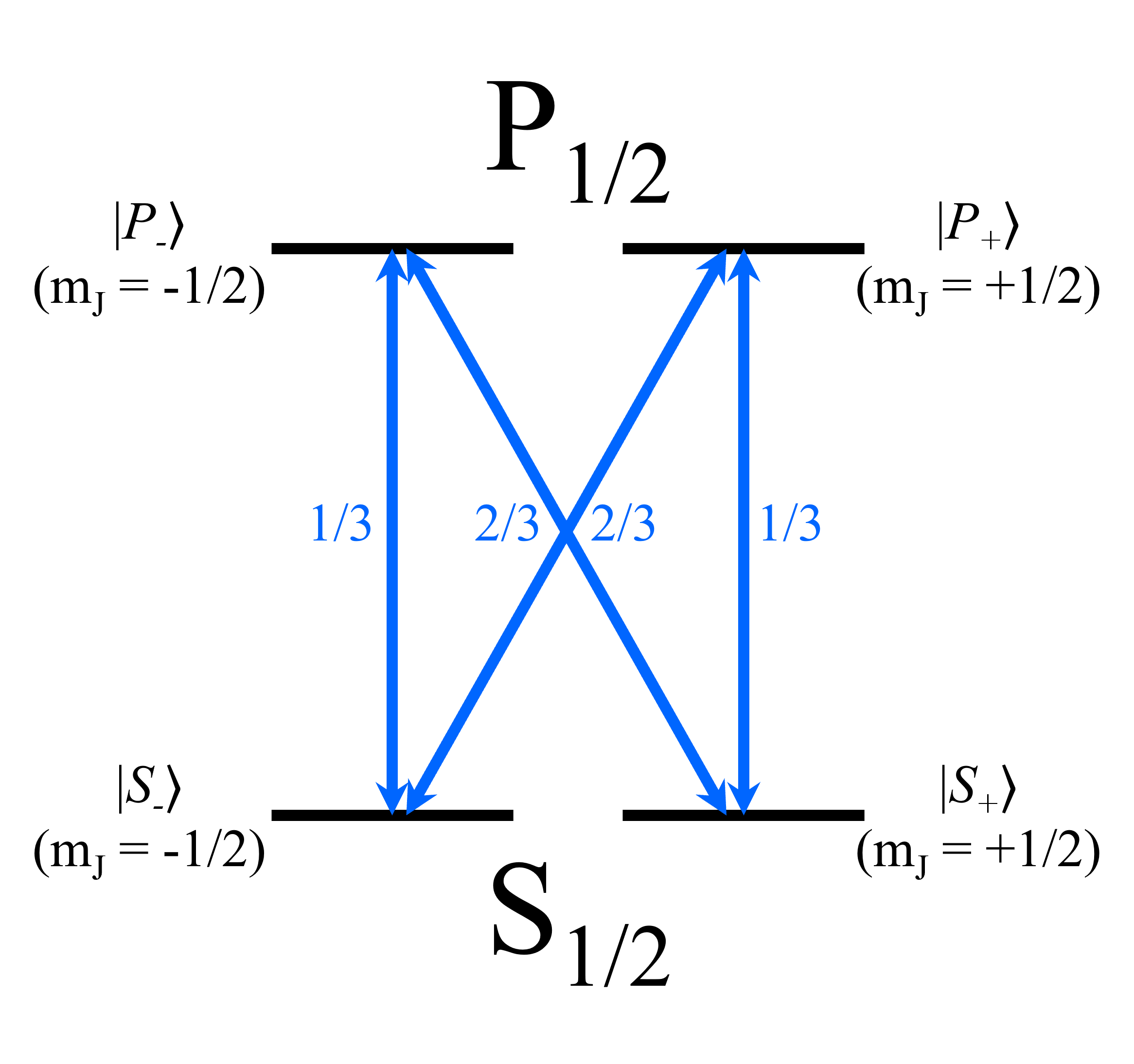}
    \caption{Partial atomic structure of ions considered in this paper consisting of a ground $S_{1/2}$ manifold and an excited $P_{1/2}$ manifold, each with two Zeeman sublevels. We label the $S$ ($P$) sublevels' state vectors with z-angular-momentum projection quantum numbers $m_J = \pm1/2$ as $|S_\pm\rangle$ ($|P_\pm\rangle$). The possible transitions are shown as blue arrows, with the corresponding Clebsch-Gordan coefficients next to the relevant arrows. }
    \label{fig:atomic structure}
\end{figure}

Consider a nuclear-spin-0 ion with a ground $S_{1/2}$ manifold and an excited $P_{1/2}$ manifold, as shown in Figure~\ref{fig:atomic structure}. By optically pumping with appropriately chosen laser parameters, such an ion can be prepared into one of the two ground state Zeeman sublevels. Following this, a circularly polarized resonant pulse can excite the ion to the $P_{1/2}$ manifold, after which the ion will decay back to the ground manifold and emit a photon whose polarization is entangled with the ion's electronic state. Performing this in two systems simultaneously, the emitted photons can be entangled and measured, resulting in REG between the systems. In this procedure, the preparation step is slow (typically $\approx$ 1 $\mu$s~\cite{Stephenson2020}) compared to the $P_{1/2}$ lifetime (typically $\approx$ 10 ns~\cite{NIST}).

\begin{figure}
    \centering
    \includegraphics[width=0.75\linewidth]{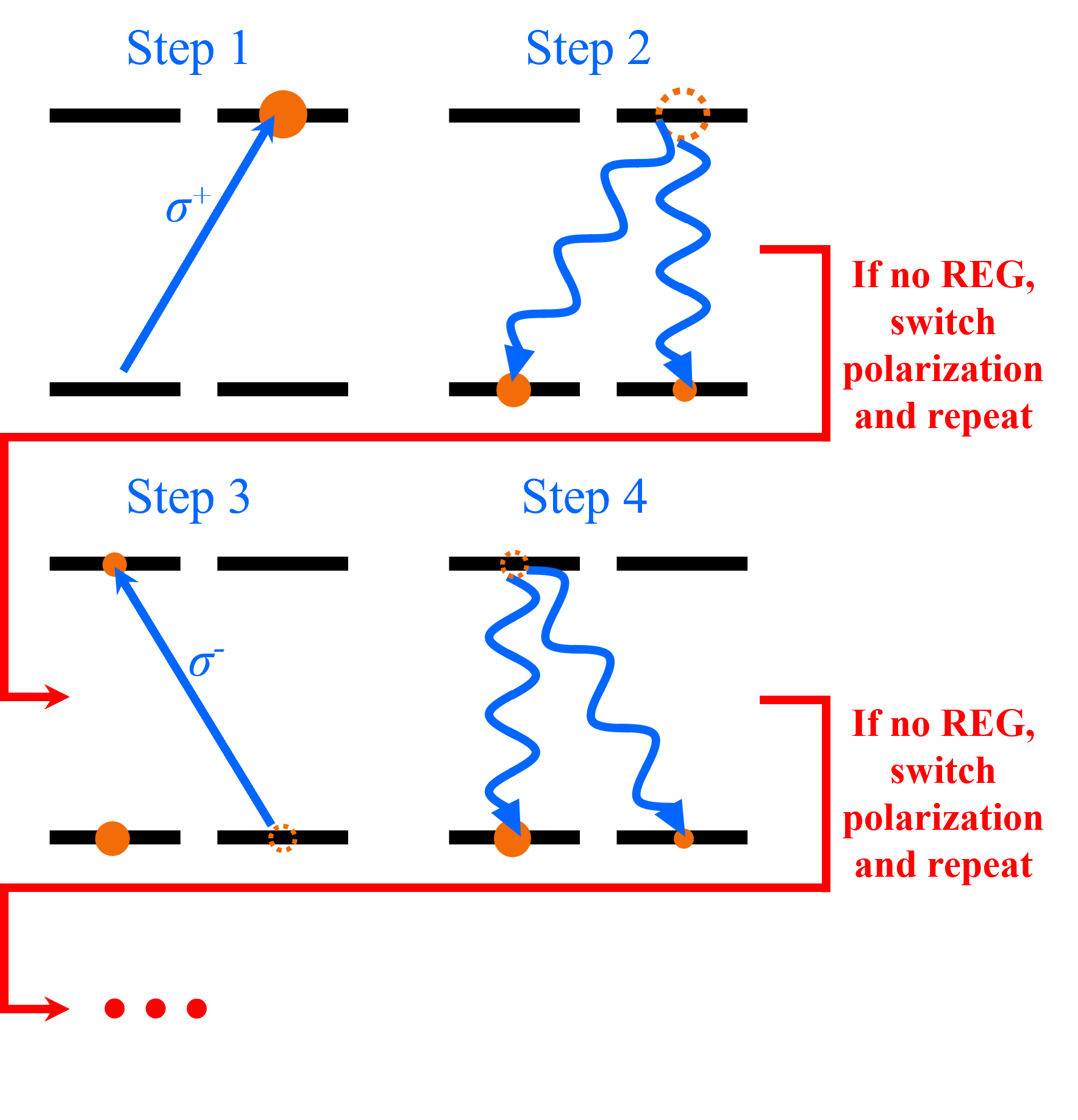}
    \caption{Schematic depiction of the electron juggling scheme. For clarity, we assume the ion is excited by $\sigma^+$ light in Step 1. The ion's internal population distribution at each step is characterized by filled orange circles, whose radii are roughly proportional to the probability the ion occupies that energy level. Dotted circles correspond to population flux out of the associated energy level. Photon detection takes place during steps 2, 4, ... etc. If photons successfully herald REG on one of these steps, the procedure is halted.}
    \label{fig:electron juggling}
\end{figure}

The electron juggling method, schematically shown in Figure~\ref{fig:electron juggling}, skips the slow optical pumping step. Instead, each REG attempt consists of a single circularly polarized excitation pulse of length $\approx$ 1\,ns or less, driving a $\sigma^+$ or $\sigma^-$ transition. After each pulse, a REG event may be heralded by detection of single photons. If the attempt fails, one then applies another circularly polarized pulse with the opposite handedness, repeating until REG is heralded. By minimizing state-preparation, this method can reduce the attempt time by about a factor of 10.

The drawbacks of this approach are: 1) Reduced probability of successful excitation, since the ion may have decayed in the previous attempt to a sublevel not addressed by the next excitation pulse, and 2) Increased sensitivity to excitation pulse polarization impurity. At a cost in REG rate, one could also apply $n$ ``state preparation" excitation pulses of one circular polarization, waiting for the ion to decay after each excitation, then apply one final excitation pulse of the opposite circular polarization. The first $n$ pulses will prepare the ion in one of the $S_{1/2}$ sublevels, and the final pulse will excite the population in that state - somewhat akin to the current slow state-preparation method typically used. After the final excitation, one can detect photons to herald REG. Using multiple pulses per attempt slows down REG, but decreases the sensitivity to polarization impurities. Furthermore, changing the pulse polarization less frequently helps prevent exciting the ion to higher energy manifolds. In this work we choose to analyze the fastest possible method where the pulse polarization alternates every pulse and show that experimentally realistic polarization purities yield high fidelity entanglement despite the scheme's higher sensitivity to this error.

Throughout the rest of this paper, we consider the application of electron juggling to the most commonly used ions for networking and distributed quantum computing at the time of writing ($^{40}$Ca$^+$, $^{88}$Sr$^+$, $^{138}$Ba$^+$, and $^{174}$Yb$^+$) in our analysis ~\cite{OReilly2024,Stephenson2020,Saha2024,Moehring2007,Maunz2009,Slodicka2013,Casabone2013,Hucul2015,Krutyanskiy2023}. $^{24}$Mg$^+$ is also included due to its favorable atomic structure which could potentially achieve high REG rates. We note that the method can be applied more generally to other physical qubits, such as neutral atoms, when they have a similarly simple energy level structure.
Furthermore, we omit isotopes with non-zero nuclear spin from our analysis due to their added complexity.

\subsection{\label{sec:repumping} Additional Repumping}
For trapped ions without low-lying D-states, such as $^{24}$Mg$^+$, the above scheme works just as described in section \ref{sec:electron juggling}, requiring only the pulsed laser to generate photons. However, in other species, such as alkaline earth ions, the electron can decay into the metastable $D_{3/2}$ manifold where natural lifetimes are as high as, \textit{e.g.}, 80 seconds in $^{138}$Ba$^+$~\cite{Sahoo2006}. If this manifold is not repumped, population will accumulate in the metastable manifold until it naturally decays. Therefore, it is necessary in such ions to repump out of the $D_{3/2}$ manifold. There are three options for repumping that can ensure a high attempt rate is maintained. First, one could continuously repump the $D_{3/2}$ manifold through the $P_{1/2}$ manifold. This approach has the advantage that it only requires one additional laser and tolerates slow modulation. To quickly repump out of the metastable state, the transition must be strongly driven, but driving too strongly can pull excited state population out of the $P_{1/2}$ manifold, which will also reduce the REG rate. The second option is to repump using \textit{shaped} pulses. By pulsing the drive to only repump when excitation and photon collection are not occurring, the photon shape is unaffected. If the pulses are kept sufficiently short and strong, and timed well out of sync with the excitation, the REG can continue unimpeded. The downside of this approach is that it requires fast modulation and strong pulses in the repump beam. The third option is to continuously repump via the $P_{3/2}$ transition. This approach combines the upsides of the two above approaches while requiring two additional lasers beyond the pulsed laser. The $D_{3/2}$ manifold can be continuously repumped through the $P_{3/2}$ manifold, and any photons produced by decays from the $P_{3/2}$ manifold can be filtered out in the imaging system. However, the ion can also decay into the $D_{5/2}$ manifold, so a second laser is required to repump out of this state.

Throughout the rest of this paper, we analyze the electron juggling scheme assuming the third option: repumping through $P_{3/2}$. This option presents the best application of electron juggling since it allows for continuous repumping without sacrificing fidelity (due to double excitations) or slowing the REG rate. The only exception to this is in the analysis of $^{174}$Yb$^+$, where we assume repumping through the bracket state $^3[3/2]_{1/2}$, as this avoids issues with repumping through $P_{1/2}$ while also avoiding population in the $D_{5/2}$ which can decay to the very long lived $F_{7/2}$ manifold \cite{Roberts2000, Biemont1998, Taylor1997}.

\section{Model}\label{sec:model}

The pulsed excitation process described in figure \ref{fig:electron juggling} is modeled by a quantum channel, which allows for polarization impurities; see Appendix~\ref{app:birefringence} for its detailed form and derivation. We model the repumping dynamics with a Lindbladian master equation,

\begin{equation}\label{Eqn:lin}
    \dot{\hat{\rho}} = -\frac{i}{\hbar} [\hat{\rho},\hat{H}] + \frac{1}{2}\sum_{\hat{C} \in \{\hat{\Gamma}_{ge}\}}[2 \hat{C} \hat{\rho} \hat{C^{\dagger}} - \hat{\rho}  \hat{C^{\dagger}} \hat{C} -  \hat{C^{\dagger}} \hat{C} \hat{\rho}],
\end{equation} 

\noindent
where $\hat{\rho}$ is the density matrix of the electron state, $\hat{\Gamma}_{ge}$ are all of the collapse operators that dissipate the state (explicitly defined below), and the repumping Hamiltonian $\hat{H}$ is given by

\begin{equation}
    \hat{H} = \sum_{j,g,e} \Omega_{j,ge} e^{i \Delta_{j,ge} t} |g\rangle\langle e| + h.c.
\end{equation}

Here the sum is over all beams $j$, and all levels $g$ and $e$ coupled by the beams. In this Hamiltonian, $\Omega_{j,ge}$ is the Rabi coupling strength for the transition between states $|g\rangle$ and $|e\rangle$ due to the beam $j$, and $\Delta_{j,ge}$ is the detuning of beam $j$ relative to the transition $g \leftrightarrow e$ (considered negative when the frequency of the beam is less than the frequency of the transition).

The collapse operators, $\hat{\Gamma}_{ge}$, include only terms for decay from excited states. The decay operators are given by~\cite{Szwer2009} 

\begin{equation}
    \hat{\Gamma}_{ge} = \sqrt{A_{ge}(2 J_e + 1)} \begin{pmatrix}
  J_g & 1 & J_e \\
  m_g & m_g-m_e & -m_e
\end{pmatrix} |g\rangle \langle e|,
\end{equation}
where $A_{ge}$ is the Einstein A coefficient for decays from the manifold containing $e$ to the manifold containing $g$, and $J_i$,  $m_i$ are the quantum numbers of the total spin-orbital angular momentum and its projection for level $i$.

In a standard Hong-Ou-Mandel (HOM) interferometer setup for two-photon detection, there are two ion-ion Bell states which may be heralded, depending on the detector click pattern~\cite{HanneganThesis}:

\begin{align}
    |\Psi^+\rangle  & = \frac{1}{\sqrt{2}}\left( |S_+S_-\rangle + |S_-S_+\rangle \right) \\
    |\Psi^-\rangle  & = \frac{1}{\sqrt{2}}\left( |S_+S_-\rangle - |S_-S_+\rangle \right).
\end{align}

The performance of the electron juggling scheme depends on the overlap of its actual output states compared with these ideal output states. Our model and simulation only consider errors due to decays from the ``wrong" $P_{1/2}$ sublevel (by ``wrong" sublevel, we mean the $P_{1/2}$ sublevel to which we do not intend to excite to, so the wrong sublevel for $\sigma^+$ ($\sigma^-$) transitions is $|P_-\rangle$ ($|P_+\rangle$)). If a photon decays from the wrong $P_{1/2}$ sublevel, the spin-polarization entanglement will be flipped. Because the Clebsch-Gordan coefficients for both possible $\sigma$ and $\pi$ transitions, respectively, have the same magnitude, a decay from the wrong sublevel amounts to a bit-flip error on the respective ion (up to a global phase). So, supposing there was a decay from the wrong sublevel in the first ion, the resulting output states for a given click-pattern would be

\begin{align}
    |\Psi^+\rangle  & \rightarrow |\Phi^+\rangle = \frac{1}{\sqrt{2}}\left( |S_-S_-\rangle + |S_+S_+\rangle \right) \\
    |\Psi^-\rangle  & \rightarrow |\Phi^-\rangle = \frac{1}{\sqrt{2}}\left( |S_-S_-\rangle - |S_+S_+\rangle \right).
\end{align}

Therefore, if one ion decays from the wrong sublevel and the detector click pattern heralds REG, the actual output state is orthogonal to the ideal output state, yielding a fidelity of zero. If, however, \textit{both} ions decay from the wrong sublevels and REG is heralded, the resulting two bit-flip errors return the same output state as the ideal. Thus, the error of the scheme is just the probability of a single decay from the wrong sublevel, meaning the fidelity is given by

\begin{equation}\label{eqn:fidelity}
    F = 1 - \frac{2 p_r p_w}{(p_r+p_w)^2},
\end{equation}
where $p_r$ and $p_w$ are the probabilities of decay from the right and wrong sublevels respectively.


The model is simulated in shots. A shot in this simulation is defined as the application of one excitation pulse (represented by a quantum channel, see Appendix~\ref{app:birefringence} for derivation) followed by the Lindbladian evolution in Eq. \ref{Eqn:lin} for repumping. On the next shot, the handedness of the excitation pulse polarization is switched. This process is repeated for many shots. For each shot of the simulation, the probability of generating a photon is computed from the density matrices representing the system just after excitation and after idling. The simulation is run for hundreds of shots to ensure that the system reaches a steady state, and then the average photon probability per shot is calculated. The photon statistics in each shot are collected and used to calculate an average rate and fidelity for the scheme.

For direct comparison with the current REG rate record in trapped ions (250 s$^{-1}$, \cite{OReilly2024}), we use the same total photon detection efficiency of 2.5\% in our simulated data. In all ion species but $^{174}$Yb$^+$, we assume each repump beam has a 30\,$\mu$m waist and 10\,$\mu$W of power. For $^{174}$Yb$^+$, we assume 200\,$\mu$W beam power with a 30\,$\mu$m waist, since more power is required to get comparable repumping rates through the $^3[3/2]_{1/2}$ bracket state. We assume the $D_{5/2}$ manifold is repumped using two beams, one which drives only $\sigma^+$ transitions and has a 10\,MHz detuning, the other which drives only $\sigma^-$ transitions and has a -10\,MHz detuning. Applying two beams detuned from one another helps avoid dark resonances which could decrease the repumping rate via coherent population trapping. However, one could instead repump with a single beam if dark resonances are disrupted some other way, such as polarization or frequency dithering~\cite{Berkeland2002}.

Next we add polarization impurities and include electronic/photonic signal propagation and processing latencies into the above model (see Appendix~\ref{app:birefringence} for details). Here, polarization impurities are assumed to be caused by birefringence~\cite{Keselman2011, Williams2013}, which can result in some opposite circular polarization in the beam. This is particularly a problem for our electron juggling scheme because of the much higher probability of the ion having population in both $S_{1/2}$ sublevels for each excitation compared with a scheme that pre-prepares a specific state with high fidelity. Latencies are important to consider in the electron juggling scheme since the attempt time can be pushed towards the atomic limit, at which point delays on the order of 10\,ns appreciably alter the REG rate achievable.

\section{Results and Discussion}\label{sec:results}
\subsection{Ideal Simulation}\label{sec:ideal}

\begin{figure}
    \centering
    \includegraphics[width=1\linewidth]{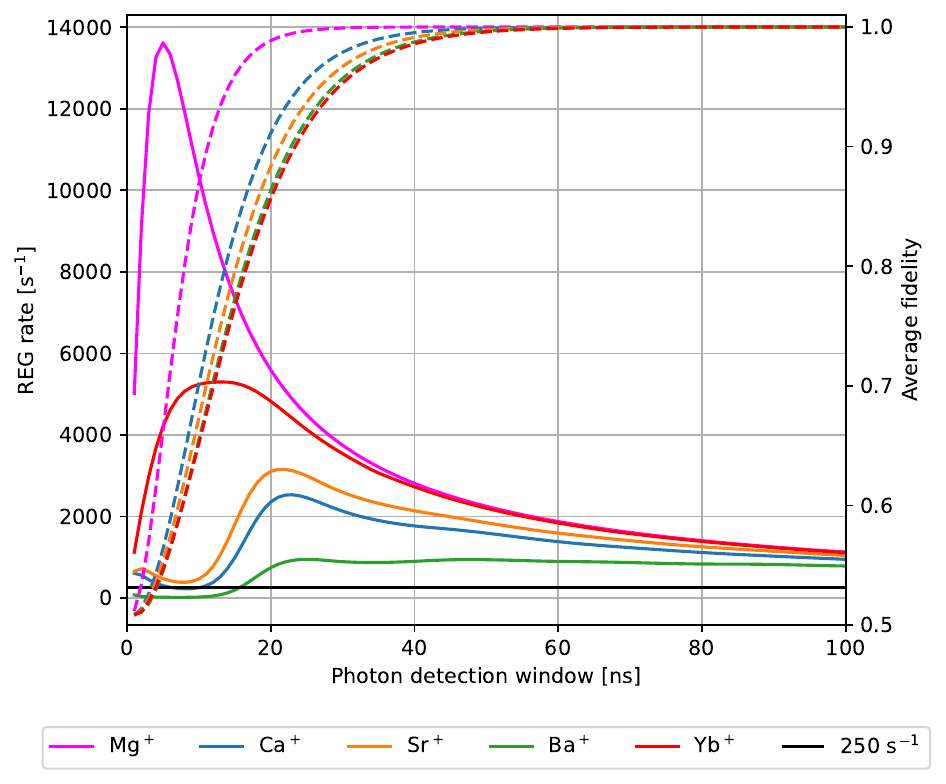}
    \caption{Results of the electron juggling simulation, assuming no latencies and no polarization impurities. The REG rate (solid curves) in Bell pairs per second and fidelity (dotted curves) are plotted against the photon detection window width. The current trapped ion REG rate record, 250 s$^{-1}$ \cite{OReilly2024} is represented by the black horizontal line. The fidelity sharply drops at small detection windows, despite assuming perfectly polarization-pure excitation pulses. This is due to residual population in the ``wrong" $P_{1/2}$ sublevel for very rapid attempts.}
    \label{fig:no latency}
\end{figure}

Figure \,\ref{fig:no latency} shows the simulation results for the idealized case of perfect polarization purity and zero latencies. In this idealized case, the electron juggling scheme enables a REG rate of over $1000\,s^{-1}$ in each species except in $^{138}$Ba$^+$ where the scheme still manages to increase the rate above the current record of $250\,s^{-1}$ by over a factor of 3.

In $^{24}$Mg$^+$, and $^{174}$Yb$^+$, the REG rate as a function of the photon detection window essentially traces out the photon's ``shape" in time; but for the alkaline earth ions (Ca$^+$, Sr$^+$, and Ba$^+$), the REG rate decreases initially before rising to its maximum. This is due to the extremely high photon generation rate in this regime. With no conventional state preparation time, the photon generation can be faster than the repumping dynamics, resulting in an appreciable chance of the alkaline earth ions falling into the $D_{3/2}$ manifold. This does not occur in $^{24}$Mg$^+$ because it has no low-lying $D$ manifold, and the effect is not noticeable in $^{174}$Yb$^+$ because its $D_{3/2}$ branching ratio is very small (0.5\%) \cite{Olmschenk2007}.

The average Bell pair fidelity is also plotted in Figure~\ref{fig:no latency}. For very small photon detection windows, the fidelity approaches 50\%. The reason for this is that, in this regime, the ion is being excited faster than the lifetime of the $P_{1/2}$ manifold. After many such excitation pulses, the ion's state is a roughly even mixture of the two $S_{1/2}$ sublevels and the two $P_{1/2}$ sublevels. Therefore, the ion has an approximately equal chance to emit a photon from each $P_{1/2}$ sublevel, and, in the language of Equation~\ref{eqn:fidelity}, $p_r \approx p_w \approx 1/2$, thus the fidelity is also $\approx 1/2$. As the photon detection window increases past the $P_{1/2}$ lifetime in Figure~\ref{fig:no latency}, the electron juggling scheme approaches its maximum achievable fidelity, which is 1 in this case since we have assumed there are no polarization impurities present.

\subsection{Impurities and Latencies Simulation}\label{sec:complexities}

\begin{figure}
    \centering
    \includegraphics[width=1\linewidth]{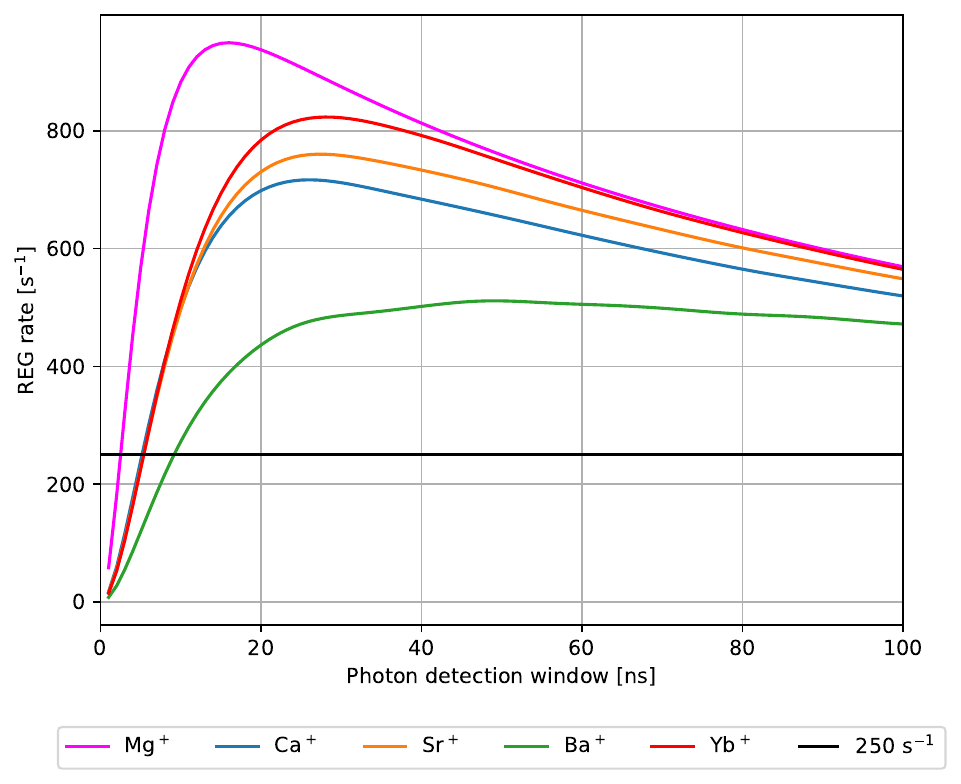}
    \caption{Results of the electron juggling simulation including polarization impurity and attempt latencies. The REG rate in Bell pairs per second is plotted against the photon detection window width (note the total attempt time will be 100\,ns plus this, since we assume 100\,ns of latencies). The current trapped ion REG rate record of 250 s$^{-1}$ \cite{OReilly2024} is represented by the black horizontal line. All species can achieve at least three times this with electron juggling.}
    \label{fig:simulation results}
\end{figure}

Figure~\ref{fig:simulation results} shows the results of a simulation which considers potentially imperfect polarization purity as well as latency in the time between a detection window and the following excitation pulse. We assume polarization impurity comes from a birefringent medium with a retarding parameter $\beta$ normally distributed with a standard deviation $\sigma_\beta = \sqrt{0.02}$; this value corresponds to a polarization intensity impurity maintained to within about 1\% of the total intensity. We also assume an attempt latency of 100\,ns~\cite{Stephenson2020}. Compared to the ideal case of no impurities and zero latencies, the peak REG rate is notably reduced for each ion species. Nonetheless, the scheme significantly improves the rate compared to the standard optical pumping approach, and as control system latencies are improved in the future, the electron juggling scheme will allow photonic interconnects to approach the atomic physics limit of REG rate.

The average fidelities are not plotted in Figure~\ref{fig:simulation results}, since in this case they are nearly constant within a narrow range between 0.984 and 0.986, depending on the ion (see Appendix~\ref{app:birefringence} for a model of this error); the rising fidelity behavior in Figure~\ref{fig:no latency} does not appear here because the attempt time is fixed to be at least 100\,ns (due to the assumed latencies). After 100\,ns, there is a negligible amount of residual $P_{1/2}$ population to decay in subsequent attempts to cause such errors (as all ions considered have a $P_{1/2}$ lifetime $\lesssim 8$\,ns~\cite{NIST}). Therefore, the fidelity of the electron juggling scheme in this simulation, is roughly constant, with its infidelity coming from the assumed polarization impurities. 

\section{Conclusion}

We have presented a method for pushing the photonic interconnect attempt rate towards the fundamental limit set by the ion's natural lifetime in the excited state. Based on our analysis, this electron juggling scheme could achieve significant rate increases over the standard optical pumping approach, even assuming realistic latencies. A summary of the simulation results is presented in Table \ref{tab:results summary} for the highest simulated REG rate where a fidelity $\geq 0.97$ (the current trapped ion REG fidelity record~\cite{Saha2024}) is also achieved. The potential gains are even higher if these latencies can be suppressed, \textit{e.g.}, by using shorter cables and optical fibers and by using faster control hardware and software for photon detection; as such latencies are improved, the electron juggling scheme has the potential to generate greater than 1,000\,s$^{-1}$ REG rates and approach the atomic physics limit of photonic interconnect speed in trapped-ion photon interconnects.

\setlength{\extrarowheight}{3pt}
\begin{table}[h!]
    \centering
    \begin{tabular}{ccc}
        \toprule\toprule
        & Rate [s$^{-1}$] & \begin{tabular}{@{}c@{}} Rate [s$^{-1}$] \\ (with latencies and \\polarization impurities) \end{tabular}\\
        \hline
        $^{24}$Mg$^+$  & 6,540 & 950 \\
        $^{40}$Ca$^+$ & 2,258 & 717 \\
        $^{88}$Sr$^+$ & 2,526 & 760 \\
        $^{138}$Ba$^+$ & 941 & 511 \\
        $^{174}$Yb$^+$ & 3,146  & 823\\
        \bottomrule\bottomrule
    \end{tabular}
    \caption{Simulation results for the electron juggling scheme. The tabulated values are the maximum achievable REG rates with a Bell pair fidelity of at least 0.97. Note that the simulation only considers errors associated with erroneous photon emission events, which can be induced either by residual $P_{1/2}$ population or polarization impurities in the excitation pulses.}
    \label{tab:results summary}
\end{table}


\section{Acknowledgments}
Simulation data was taken using QuTiP~\cite{Lambert2024}. We thank Mihir Bhaskar for helpful comments.
\nocite{*}

\bibliography{biblio}

\clearpage

\appendix

\section{Derivation of Excitation Pulse Superoperator}\label{app:birefringence}

Assuming our excitation pulse is square with ideal circular polarization of handedness $h$ ($h$ can be $r$ for right-hand circular or $l$ for left-hand circular), its Hamiltonian $H$ may be written as

\begin{equation}
    H = H_{h,+} + H_{h,-},
\end{equation}
where, in the interaction picture,

\begin{equation}
    H_{h,\pm} = \frac{\Omega_{h,\pm}}{2} e^{i\Delta t} |S_\mp \rangle \langle P_\pm| + \frac{\Omega_{h,\pm}^*}{2} e^{-i\Delta t} |P_\pm \rangle \langle S_\mp|,
\end{equation}
where $\Omega_{h,\pm}$ is the Rabi frequency of the transition driven by $\sigma^\pm$ polarized light for a beam with ideal circular polarization handedness $H$; $\Delta$ is the detuning of the laser beam from the dipole transition it addresses. Note that $H_{h,+}$ denotes $\sigma^+$ drive and $H_{h,-}$ denotes $\sigma^-$ drive. Thus, in the ideal case, $H_{h,-}$ will be the zero operator when we apply a $\sigma^+$ pulse, and $H_{h,+}$ will be zero when we apply a $\sigma_-$ pulse. However, this will not be true if the pulses have polarization impurities.

Now, since we want to rapidly excite the ion, the pulse will be very powerful, close to resonance, and no longer than $\pi/\Omega_{h,\pm}$. This implies that, during excitation, $\Delta t \leq \pi \Delta/\Omega_{h,\pm} \ll 1$, and we can therefore safely neglect the time-dependent phases, giving

\begin{equation}
    H_{h,\pm} \approx \frac{\Omega_{h,\pm}}{2} |S_\mp \rangle \langle P_\pm| + \frac{\Omega_{h,\pm}^*}{2} |P_\pm \rangle \langle S_\mp|.
\end{equation}

Because this Hamiltonian is time-independent, we can directly integrate the Schr$\ddot{\text{o}}$dinger equation to find the propagator. Furthermore, since $H_{h,+}$ and $H_{h,-}$ operate on orthogonal subspaces of the overall Hilbert space, the final propagator will be the sum of the propagator for each Hamiltonian. Therefore, we find that the propagator is

\begin{equation}
    U_H = U_{h,+} + U_{h,-}
\end{equation}
where

\begin{equation}
    U_{h,\pm} = e^{-i[\theta_{h,\pm}/2 |S_\mp \rangle \langle P_\pm| + \theta_{h,\pm}^*/2 |P_\pm \rangle \langle S_\mp|]},
\end{equation}
where $\theta_{h,\pm} = \Omega_{h,\pm} t$.

Taylor expanding this operator yields

\begin{equation}
    U_{h,\pm} =\sum^\infty_{n=0} \frac{-[i(\theta_{h,\pm}/2 |S_\mp \rangle \langle P_\pm| + \theta_{h,\pm}^*/2 |P_\pm \rangle \langle S_\mp|)]^n}{n!}.
\end{equation}
Defining $e^{i\phi_\pm} \equiv \theta_{h,\pm}/|\theta_{h,\pm}|$ and writing out the first few terms of this expansion gives

\begin{align}
n=1: & -i(\theta_{h,\pm} |S_\mp \rangle \langle P_\pm | + \theta_{h,\pm}^* |P_\pm \rangle \langle S_\mp |) \nonumber \\ & =-i|\theta_{h,\pm}|(\frac{\theta_{h,\pm}}{|\theta_{h,\pm}|} |S_\mp \rangle \langle P_\pm | + \frac{\theta_{h,\pm}^*}{|\theta_{h,\pm}|} |P_\pm \rangle \langle S_\mp |) \nonumber \\ & =-i|\theta_{h,\pm}|(e^{i\phi_\pm} |S_\mp \rangle \langle P_\pm | + e^{-i\phi_\pm} |P_\pm \rangle \langle S_\mp |) \nonumber \\ & \equiv -i|\theta_{\pm}| \hat{X}_{\phi,\pm} \nonumber \\
n=2: & -\frac{1}{2!} \frac{|\theta_{h,\pm}|^2}{4} (|S_\mp \rangle \langle S_\mp | + |P_\pm| \rangle \langle P_\pm|) \nonumber \\
& \equiv -\frac{1}{2!} \frac{|\theta_{h,\pm}|^2}{4} \hat{\mathds{1}}_\pm  \nonumber \\
n=3: & -\frac{i}{3!} \frac{|\theta_{h,\pm}|^2}{4} (\theta_{h,\pm} |S_\mp \rangle \langle P_\pm | + \theta_{h,\pm}^* |P_\pm \rangle \langle S_\mp |) \nonumber \\ & = -\frac{i}{3!} \frac{|\theta_{h,\pm}|^3}{8} \hat{X}_{\phi,\pm},
\end{align}
where $\hat{\mathds{1}}_\pm$ is the direct sum of the projector onto the subspace spanned by $|S_\mp\rangle$ and $|P_\pm\rangle$ and the zero operator on the complementary subspace; similarly, $\hat{X}_{\phi,\pm}$ is the direct sum of the bit-flip operator (rotated through an angle $\phi$ in the equator of the Bloch sphere) on the subspace spanned by $|S_\mp\rangle$ and $|P_\pm\rangle$ and the zero operator on the complementary subspace.

From these first few terms, it is clear from inductive reasoning that

\begin{equation}
    U_\pm = \text{cos}(\frac{|\theta|_\pm}{2}) \hat{\mathds{1}}_\pm - i\,\text{sin}(\frac{|\theta_{h,\pm}|}{2}) \hat{X}_{\phi,\pm}.
\end{equation}
Before applying these operators to the states, we want to introduce polarization impurities. Specifically, we are interested in polarization impurities caused by birefringence, which will retard orthogonal polarization components along the principle axes of the medium, causing the handedness of our (ideally) pure circular polarization to become mixed. We can model this with the following Jones matrix

\begin{align*}
J_r &= \begin{pmatrix} \text{cos}(\alpha) & \text{sin}(\alpha) \\ -\text{sin}(\alpha) & \text{cos}(\alpha) \end{pmatrix}
      \begin{pmatrix} e^{i \beta/2} & 0 \\ 0 & e^{-i \beta/2} \end{pmatrix}
      \begin{pmatrix} \text{cos}(\alpha) & -\text{sin}(\alpha) \\ \text{sin}(\alpha) & \text{cos}(\alpha)
      \end{pmatrix}.
\end{align*}
Here, we assume a fixed coordinate system. The right and left rotation matrices, respectively, rotate us into and out of the coordinate system defined by the ``fast" and ``slow" axes of the birefringent medium, which is assumed to be an angle $\alpha$ off of our chosen coordinate system. The middle matrix acts to phase shift the fast and slow components of the beam's polarization, which creates the polarization impurity in circularly polarized beams.

The Jones vectors for $\sigma_+$ and $\sigma_-$ polarized light are, respectively,

\begin{align}
    \hat{\epsilon}_+ = \begin{pmatrix} -\frac{1}{\sqrt{2}} \\  -\frac{i}{\sqrt{2}} \end{pmatrix}
\end{align}

\begin{align}
    \hat{\epsilon}_- = \begin{pmatrix} -\frac{1}{\sqrt{2}} \\  \frac{i}{\sqrt{2}} \end{pmatrix}
\end{align}

This means that an initially pure $\sigma_+$ beam impinging on this birefrengent medium would emerge with a right-hand circular polarization component of

\begin{equation}
    c_{r,+}(\alpha,\beta) =\hat{\epsilon}_+^\dagger J_r \hat{\epsilon}_+ = \text{cos}(\beta/2),
\end{equation}
and a left-hand circular polarization component of

\begin{equation}
    c_{r,-}(\alpha,\beta) =\hat{\epsilon}_-^\dagger J_r \hat{\epsilon}_+ = i e^{-i 2 \alpha} \text{sin}(\beta/2).
\end{equation}

Similarly, for a pure beam meant to drive $\sigma_-$ transitions, it would emerge from the birefringent medium with a right hand circular component of
\begin{equation}
    c_{l,+}(\alpha,\beta) =\hat{\epsilon}_+^\dagger J_r \hat{\epsilon}_- = i e^{i 2 \alpha} \text{sin}(\beta/2),
\end{equation}
and a left-hand circular component of

\begin{equation}
    c_{l,-}(\alpha,\beta) =\hat{\epsilon}_-^\dagger J_r \hat{\epsilon}_- = \text{cos}(\beta/2).
\end{equation}
Now we can compute the effect of such birefringence on the propagator. Birefrengence will alter the $\theta$ parameters as 

\begin{equation}
    \theta_{h,\pm} \rightarrow c_{h,\pm}(\alpha,\beta) \theta_H,
\end{equation}
where $H$ is the ideal handedness of the applied pulse's polarization, $R$ or $L$; $\theta_H$ here means the pulse area for the polarization handedness $H$, which will be $\pi$ for $R$ and $L$ since they are both, ideally, perfect excitation pulses. So the total propagator will be

\begin{align}
    U_H(\alpha,\beta) = & \,  U_{h,+} + U_{h,-}\\
    = & \, \text{cos}(\frac{\pi |c_{h,+}(\alpha,\beta)|}{2}) \hat{\mathds{1}}_+ \\
    &  - i\,\text{sin}(\frac{\pi |c_{h,+}(\alpha,\beta)|}{2}) \hat{X}_{\phi(\alpha,\beta),+} \\
    & + \text{cos}(\frac{\pi |c_{h,-}(\alpha,\beta)|}{2}) \hat{\mathds{1}}_- \\
    &  - i\,\text{sin}(\frac{\pi |c_{h,-}(\alpha,\beta)|}{2}) \hat{X}_{\phi(\alpha,\beta),-}
\end{align}
Now, if the inital state of the ion is $\rho_0$, the final state will be

\begin{equation}
    \rho_f(\alpha,\beta) = U_H^\dagger \rho_0 U_H.
\end{equation}
Now we are prepared to average over the parameters of the birefringent medium. We will assume the birefringent fast and slow axes, determined by $\alpha$ are randomly selected from a uniform distribution, and that the retardance applied to the photons, determined by $\beta$, is normally distributed. Then we have

\begin{equation}
    \langle \rho_f \rangle = \frac{1}{\pi} \frac{1}{\sqrt{2 \pi} \sigma_\beta}\int_{-\infty}^{\infty}\int_0^\pi U_H^\dagger \rho_0 U_H e^{\frac{-\beta^2}{2\sigma_\beta^2}}d\alpha d\beta.
\end{equation}
In this equation, any term proportional to $\hat{X}_{\phi(\alpha,\beta),\pm}$ is identically zero, since

\begin{equation}
    \int^\pi_0e^{\pm i 2 \alpha} d\alpha = 0.
\end{equation}

Therefore, the superoperator corresponding to a circularly polarized (with handedness $H$) excitation pulse which passed through a birefringent medium before reaching the ion is

\begin{align}
\mathscr{E}(\rho_0) & = \frac{1}{\pi} \frac{1}{\sqrt{2\pi} \sigma_\beta} \int^\infty_{-\infty} \int^\pi_0 e^{\frac{-\beta^2}{2\sigma_\beta^2}} \left[ \text{cos}^2\left(\frac{\pi |c_{h,+}|}{2}\right) \hat{\mathds{1}}_+ \rho_0 \hat{\mathds{1}}_+ \right. \\
 & \quad + \text{sin}^2\left(\frac{\pi |c_{h,+}|}{2}\right) \hat{X}_{\phi(\alpha,\beta),+}\rho_0\hat{X}_{\phi(\alpha,\beta),+} \nonumber \\
& \quad + \text{cos}^2\left(\frac{\pi |c_{h,-}|}{2}\right) \hat{\mathds{1}}_- \rho_0 \hat{\mathds{1}}_- \nonumber \\
 & \quad \left. + \text{sin}^2\left(\frac{\pi |c_{h,-}|}{2}\right) \hat{X}_{\phi(\alpha,\beta),-}\rho_0\hat{X}_{\phi(\alpha,\beta),-} \right] d\alpha d\beta. \nonumber
\end{align}
The terms $\hat{X}_{\phi(\alpha,\beta),\pm}\rho_0\hat{X}_{\phi(\alpha,\beta),\pm}$ expand as

\begin{align}
    \hat{X}_{\phi(\alpha,\beta),\pm}\rho_0\hat{X}_{\phi(\alpha,\beta),\pm} = & |S_\mp \rangle \langle P_\pm|\rho_0 |P_\pm \rangle \langle S_\mp| \nonumber \\
    & + |P_\pm \rangle \langle S_\mp|\rho_0 |S_\mp \rangle \langle P_\pm| \nonumber \\
    & + e^{i 2 \phi_\pm}|S_\mp \rangle \langle P_\pm|\rho_0 |S_\mp \rangle \langle P_\pm| \nonumber \\
    & + e^{i 2 \phi_\pm}|P_\pm \rangle \langle S_\mp|\rho_0 |P_\pm \rangle \langle S_\mp|,
\end{align}
but the latter two terms, being proporitional to ${e^{i 2 \phi_\pm} \propto e^{\pm i 4 \alpha}}$, integrate to zero. So we have

\begin{align}
    \hat{X}_{\phi(\alpha,\beta),\pm}\rho_0\hat{X}_{\phi(\alpha,\beta),\pm} = & |S_\mp \rangle \langle P_\pm|\rho_0 |P_\pm \rangle \langle S_\mp| \nonumber \\
    & + |P_\pm \rangle \langle S_\mp|\rho_0 |S_\mp \rangle \langle P_\pm| \nonumber \\
    \equiv & \quad \rho_{X\pm}.
\end{align}
Since the coefficients $|c_{h,\pm}|$ depend only on $\beta$, the superoperator has no remaining $\alpha$ dependence and that integral may therefore be dropped, leaving

\begin{align}
\mathscr{E}(\rho_0) & = \frac{1}{\sqrt{2\pi} \sigma_\beta} \int^\infty_{-\infty} e^{\frac{-\beta^2}{2\sigma_\beta^2}} \left[ \text{cos}^2\left(\frac{\pi |c_{h,+}|}{2}\right) \hat{\mathds{1}}_+ \rho_0 \hat{\mathds{1}}_+ \right. \\
 & \quad + \text{sin}^2\left(\frac{\pi |c_{h,+}|}{2}\right) \rho_{X+} \nonumber \\
& \quad + \text{cos}^2\left(\frac{\pi |c_{h,-}|}{2}\right) \hat{\mathds{1}}_- \rho_0 \hat{\mathds{1}}_- \nonumber \\
 & \quad \left. + \text{sin}^2\left(\frac{\pi |c_{h,-}|}{2}\right) \rho_{X-} \right] d\beta. \nonumber
\end{align}
The superoperator is therefore determined entirely by the variance of the polarization noise $\sigma_\beta$. If this parameter is small, then we can Taylor expand the trigonometric functions to get, for an ideally right-hand circularly polarized pulse (and keeping only terms lower than fourth-order in $\beta$),

\begin{align}
    \mathscr{E}_r(\rho_0) = & \quad\frac{1}{\sqrt{2\pi} \sigma_\beta} \int^\infty_{-\infty} e^{\frac{-\beta^2}{2\sigma_\beta^2}} \left[ O(\beta^4)\hat{\mathds{1}}_+ \rho_0 \hat{\mathds{1}}_+ \right.\\
    & \quad + \left(1 + O(\beta^4) \right)\rho_{X+} \nonumber \nonumber\\
    & \quad + \left(1 - \frac{\pi^2 \beta^2}{16}  + O(\beta^4) \right) \hat{\mathds{1}}_- \rho_0 \hat{\mathds{1}}_- \nonumber\\
    & \quad \left. + \left(\frac{\pi^2 \beta^2}{16} + O(\beta^4)\right) \rho_{X-} \right] d\beta.
\end{align}
These integrals are just various moments of the normal distribution, hence

\begin{align}
    \mathscr{E}_r(\rho_0) = &  \quad O(\sigma_\beta^4)\hat{\mathds{1}}_+ \rho_0 \hat{\mathds{1}}_+ \\
    & \quad +\left(1+O(\sigma_\beta^4)\right)\rho_{X+} \\
    & \quad + \left(1 - \frac{\pi^2 \sigma_\beta^2}{16} + O(\sigma_\beta^4)\right) \hat{\mathds{1}}_- \rho_0 \hat{\mathds{1}}_- \nonumber\\
    & \quad + \left(\frac{\pi^2 \sigma_\beta^2}{16} + O(\sigma_\beta^4)\right) \rho_{X-}
\end{align}
Similarly for a left-hand circularly polarized pulse we get

\begin{align}
    \mathscr{E}_l(\rho_0) = & \frac{1}{\sqrt{2\pi} \sigma_\beta} \int^\infty_{-\infty} e^{\frac{-\beta^2}{2\sigma_\beta^2}} \left[O(\beta^4)\hat{\mathds{1}}_- \rho_0 \hat{\mathds{1}}_- \right.\\
    & \quad + \left(1 + O(\beta^4) \right)\rho_{X-} \nonumber \nonumber\\
    & \quad + \left(1 - \frac{\pi^2 \beta^2}{16}  + O(\beta^4) \right) \hat{\mathds{1}}_+ \rho_0 \hat{\mathds{1}}_+ \nonumber\\
    & \quad \left. + \left(\frac{\pi^2 \beta^2}{16} + O(\beta^4)\right) \rho_{X+} \right] d\beta.
\end{align}
And integrating,

\begin{align}
    \mathscr{E}_l(\rho_0) = & \quad  O(\sigma_\beta^4)\hat{\mathds{1}}_- \rho_0 \hat{\mathds{1}}_- \\
    & \quad +\left(1 + O(\sigma_\beta^4) \right)\rho_{X-} \nonumber \nonumber\\
    & \quad + \left(1 - \frac{\pi^2 \sigma_\beta^2}{16}  + O(\sigma_\beta^4) \right) \hat{\mathds{1}}_+ \rho_0 \hat{\mathds{1}}_+ \nonumber\\
    & \quad + \left(\frac{\pi^2 \sigma_\beta^2}{16} + O(\sigma_\beta^4)\right) \rho_{X+}.
\end{align}

Therefore, conditional on the ion initially being in the ``wrong" $S_{1/2}$ sublevel, birefringence causes a ``bad" excitation probability of approximately $\pi^2\sigma_\beta^2/16$. For a standard deviation corresponding to impurities of polarization intensity of $\pm0.5\%$, $\sigma_\beta \approx \sqrt{0.02}$, so the conditional probability of a bad excitation is $\sim 1.23\%$. Based on the simulations of Section~\ref{sec:results}, the ions eventually reach a ``steady state" where the probability of a ``good" excitation is about 60\%. Using Equation~\ref{eqn:fidelity}, with $p_r = 0.6$ and $p_w = 0.4 \times 0.0123$, we would expect an average fidelity of $0.986$. This is close to the output of the simulation with polarization impurities and latencies, which yields fidelities between 0.984 and 0.986 depending on the ion. The deviation from the estimate here is likely due to the excitation probability varying depending on repumping performance for each ion and the polarization impurity.

\end{document}